\documentclass[%
 reprint,
 amsmath,amssymb,
 aps,
pra,
]{revtex4-2}

\usepackage{graphicx}
\usepackage{dcolumn}
\usepackage[dvipsnames]{xcolor}
\usepackage{bm}
\usepackage{comment}
\usepackage{verbatim}

\usepackage[normalem]{ulem}
\usepackage{xcolor}

\begin{document}

\preprint{APS/123-QED}

\title{Metropolitan entanglement distribution between an atom and a near-visible photon}

\author{Maya Büki$^{1,3}$}
\thanks{These authors contributed equally to this work}
\author{Pooja Malik$^{2,3}$}
\thanks{These authors contributed equally to this work}
\author{Florian Fertig$^{2,3}$}
\author{Tobias Frank$^{1,3}$}
\author{Marvin Scholz$^{1, 3}$}
\author{Tommy Block$^{2,3}$}
\author{Gianvito Chiarella$^{1,3}$}
\author{Yiru Zhou$^{2,3}$}
\author{Emanuele Distante$^{1,3}$}
\email{Present address: LENS - European Laboratory for Non-linear Spectroscopy \& 
Dipartimento di Fisica e Astronomia, Università degli Studi di Firenze, Via Sansone 1, 50019 Sesto Fiorentino (Firenze), Italy}
\author{Pau Farrera$^{1,3}$}
\email{Corresponding author: pau.farrera@mpq.mpg.de}
\author{Gerhard Rempe$^{1,3}$}
\author{Harald Weinfurter$^{2,3}$}
\affiliation{$^{1}$Max-Planck-Institut für Quantenoptik, Hans-Kopfermann-Straße 1, 85748 Garching, Germany}
\affiliation{$^{2}$Fakultät für Physik, Ludwig-Maxmilians-Universität München, Schellingstraße 4, 80799 München, Germany}
\affiliation{$^{3}$Munich Center for Quantum Science and Technology, Schellingstraße. 4, 80799 München, Germany}


\begin{abstract}

\noindent Entanglement distribution is the overarching purpose of quantum networks. While communication over long distances can use deployed fiber infrastructure, it requires photons in the telecom band. However, advanced quantum network nodes do not operate at such wavelengths. Here we overcome this limitation with two tailor-made low-noise quantum-frequency converters to distribute entanglement between a single atom and a resonant photon over 14\,km line-of-sight via 24\,km of a deployed commercial fiber. The photon at wavelength $780\,\mathrm{nm}$ is first entangled with the atom, then converted to the telecom S-band, and finally back-converted after propagation through the fiber. This link enables a photon transfer efficiency of 1.7\% while affecting the atom-photon entanglement fidelity by less than 1\%.
This brings integration of atomic quantum nodes with existing long-distance fiber networks into reach, enabling novel applications in quantum information processing.

\end{abstract}

\maketitle

\noindent Long-distance quantum networks offer huge potential for communicating and processing quantum information~\cite{Cirac1997, Kimble2008, Wehner2018}. Providing distributed entanglement, they form the basis for a number of quantum-information applications such as efficient quantum communication with quantum repeaters~\cite{Briegel1998, Azuma2023}, scalable quantum computation by connecting quantum processors at remote locations to larger units~\cite{Main2025}, quantum sensing~\cite{Zhang2021, Nichol2022, Novikov2025}, or long baseline telescopy~\cite{Gottesman2012}. Realizing these capabilities inevitably requires quantum networks to be modular, as components need to be combined and individually optimized for distinct tasks such as photon transmission, entanglement storage, and quantum processing~\cite{Bussieres2013, Wei2022}. Experiments distributing entanglement between a node and a photonic qubit or even between two quantum nodes were demonstrated in the laboratory~\cite{Blinov2004, Volz2006, Moehring2007, Wilk2007, Ritter2012}, later with long fibers on spools~\cite{Chang2019, VanLeent2020, Zhou2024, Krutyanskiy2024, Cui2025,Wang2026}, more recently also in implementations with real-world, deployed fibers~\cite{Rakonjac2023, Knaut2024, Kucera2024}, and even between distant nodes~\cite{VanLeent2022, Liu2024, Stolk2024, Luo2022}. 
Yet, these proof-of-concept demonstrations clearly show further challenges towards an efficient, high fidelity quantum link.

First, all operations and manipulations of the quantum systems together with communication between nodes have to be synchronized, and performed within the coherence times of the systems. Second, while existing fiber networks offer great opportunity to distribute light, mechanical and thermal drifts require automated adjustment, particularly when using polarization qubits. The main challenge, however, arises from the fact that low transmission loss is ensured only for telecom wavelengths. To this end, quantum frequency conversion (QFC) transferring the wavelength from the visible or near-visible to the telecom wavelengths without altering the quantum state has been developed~\cite{Tanzilli2005, Ramelow2012, Albrecht2014, Ikuta2018}. Unidirectional conversion suffices for implementations that employ  entanglement swapping at an additional central node. However, a bidirectional solution is required for general-purpose quantum networks, as almost all light-matter interfaces and quantum network elements such as processing nodes~\cite{Bussieres2013}, nondestructive photonic qubit detectors~\cite{Niemietz2021}, and also free space links~\cite{Liao2017}, operate in the visible or near-visible region, well outside the telecom windows. 
Therefore, a modular and widely tunable QFC from the visible to telecom bands \emph{and} vice-versa is required to connect future quantum network nodes, possibly even with quantum systems operating at different wavelengths~\cite{Maring2017, Siverns2019}. Yet, the quality of the quantum link is easily deteriorated as attenuation along the link and additional noise from the back conversion stage degrade the signal-to-noise ratio (SNR) and thus the observed entanglement. Therefore, high-fidelity and efficient bidirectional QFC for long-distance quantum communication remains a key challenge.

Here, we demonstrate a metropolitan quantum network link capable of distributing entanglement between a single atom and a near-visible photon over a 24\,km long deployed fiber. Such single atoms have demonstrated many quantum networking advantages including long coherence time, longer than the communication time~\cite{Korber2018, Zhou2024}, unique entanglement capabilities~\cite{Thomas2022, Thomas2024, Chiarella2025}, and powerful schemes for quantum state processing~\cite{Reiserer2015, Bluvstein2024, Canteri2025}, making them one of the most advanced platforms for quantum nodes~\cite{Covey2023}. We leverage bidirectional quantum frequency conversion to interface the atomic qubit - Rubidium ($^{87}$Rb) emitting at $\lambda=$ 780\,nm -  with telecommunication infrastructure (the S-band at 1514\,nm) and subsequent back-conversion to 780\,nm. We maintain a high signal-to-noise ratio through a careful selection of the telecommunication wavelength, the application of narrow-band filtering, and a thorough optimization of the QFC process. This allows us to achieve an atom-photon entanglement fidelity $\mathcal{F}>85\%$ at both wavelengths, showing little reduction from the generated entangled state at 0\,km. We characterize the preservation of the entanglement before and after the second frequency conversion, and illustrate its potential in the context of long-distance quantum networks as well as of connecting different platforms.

\begin{figure}
    \includegraphics[width=0.95\linewidth]{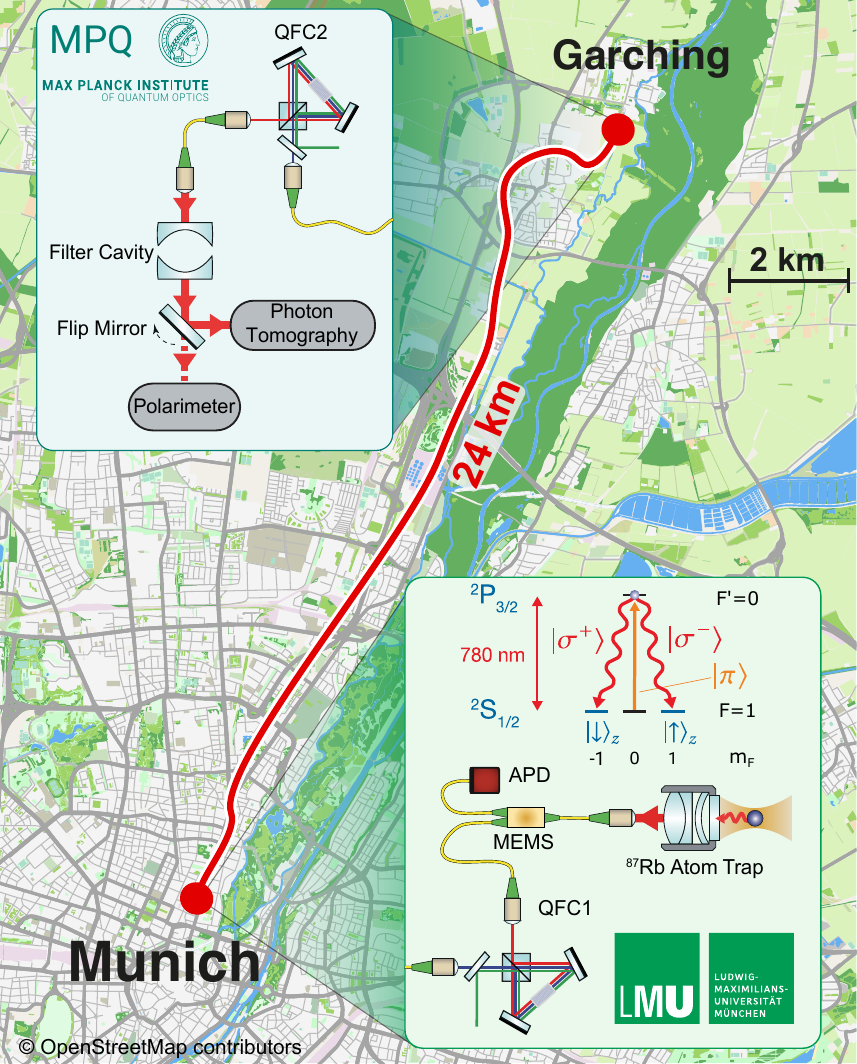}
    \caption{Metropolitan quantum link. Two laboratories in the greater Munich area are connected via a 24\,km long fiber in order to distribute atom-photon entanglement. The Rb atom at LMU (node 1) is entangled with a photonic qubit (see level scheme) which is subsequently converted to a telecom wavelength at QFC1, propagates 24\,km of fiber, is back-converted to 780\,nm at QFC2, and after spectral filtering finally measured in a photon projection measurement at MPQ (node 2). After communicating to LMU the arrival of the photon at MPQ, the atomic state is measured via florescence readout using a micro-electro-mechanical systems (MEMS) switch and an avalanche photodetector (APD). }
    \label{fig:setup}
\end{figure}

Our metropolitan fiber link connects two quantum optics laboratories separated by 14\,km line-of-sight. One is located at Ludwig-Maximilians-University (LMU) in downtown Munich, and the other one at the Max-Planck-Institute of Quantum Optics (MPQ) in the city of \mbox{Garching}, north of Munich (see Fig.\,\ref{fig:setup}). The laboratory at LMU contains a single-atom quantum node capable of generating atom-photon entanglement, and is equipped with a 780\,nm-to-1514\,nm quantum frequency converter (QFC1) \cite{Ikuta2018,VanLeent2020}. The laboratory at MPQ contains a 1514\,nm-to-780\,nm converter (QFC2) and a tomography setup to characterize the polarization state of the photon. Two 24\,km long fibers deployed in the Munich metropolitan area (provided by the telecommunication company M-Net) connect the two laboratories. One fiber carries the photon qubits, while the other transmits classical reference and synchronization signals, including the distributed pump light. The fiber used for single photons has a total attenuation of 8.41\,dB (i.e. on average 0.35\,dB/km) and polarization dependent losses of $0.109 \pm 0.019$\,dB.

The experiment starts with trapping and cooling the single atom at LMU ($^{87}\mathrm{Rb}$), and performing optical pumping into the state $\left|5^2S_{1/2}, F=1,m_F=0\right>$. After exciting the atom to the $\left|5^2P_{3/2}, F=0\right>$ level, it emits a single photon whose polarization state (after being coupled into a single mode fiber) is maximally entangled with the atom's internal state~\cite{Volz2006}. 
\begin{equation}
\left|\Psi\right\rangle = \frac{1}{\sqrt{2}}\left(\left|L,\downarrow_z\right\rangle+\left|R,\uparrow_z\right\rangle\right)
\label{eq:bellstate}
\end{equation}
Here $\left|L\right\rangle$ and $\left|R\right\rangle$ denote the left and right circular polarization of the photon, and $\left|\downarrow_z\right\rangle$ and $\left|\uparrow_z\right\rangle$ denote states $\left|5^2S_{1/2}, F=1,m_F=-1\right>$ and $\left|5^2S_{1/2}, F=1,m_F=+1\right>$, respectively. The wavelength of the photon at 780\,nm is subsequently converted to 1514\,nm (through polarization preserving difference frequency generation with a pump laser at 1610\,nm) in order to enable propagation through the metropolitan fiber with low losses~\cite{VanLeent2020}. At the same time the atomic qubit is mapped to a magnetic-field-insensitive superposition state with the help of Raman-transfer laser beams in order to extend the coherence time of the atomic qubit~\cite{Zhou2024}. After propagating through the 24\,km of deployed fiber, the photonic qubit arrives at MPQ, where its frequency is back-converted to 780\,nm (through sum frequency generation with a second pump laser at 1610\,nm), and subsequently measured. The overall entanglement distribution efficiency of the experiment is $3.1\times 10^{-5}$, which includes 1\% efficiency of collecting and coupling the emitted photon into the fiber, the full link efficiency of 1.7\% (comprising 42\% and 37\% QFC efficiency in the down- and up-conversion stages, respectively, and 11\% transmission through the metropolitan fiber), and 18\% transmission through additional optics, filtering and detection components. The detection of the photon at MPQ is communicated to LMU, where the atomic qubit is then mapped back to the original superposition state. The last step in the experimental sequence involves measuring the atomic qubit with the help of a state-dependent photoionization process and fluorescence atomic state detection (described in further detail in~\cite{suppMat}). The measurement results from both the photonic and atomic qubits states are subsequently analyzed in order to characterize the entangled state.

Several aspects are carefully considered in order to preserve the photonic qubit state all the way from its generation at LMU to its analysis at MPQ. Qubit rotations and mixing with noise photons can occur in the QFC stages and along the long fiber link, and different strategies to mitigate these effects are implemented. For the metropolitan fiber, qubit polarization rotations are the main effect to consider. Thanks to its underground deployment, the fiber is protected from rapid temperature and stress-induced birefringence fluctuations that typically occur in an open-air environment~\cite{Kucera2024,Bersin2024}. However, despite the good passive stability, long-term drifts are observable, causing unitary photon qubit rotations that would impact the results of long measurement runs~\cite{suppMat}. In order to correct for these slow drifts, the experiment is automatically interrupted every 8 minutes to measure the polarization state of a classical light beam (at the atomic emission wavelength) that propagates through the whole communication link. The results of these measurements are used by a gradient-descent algorithm that corrects drifts in the fiber birefringence by acting on a piezo-based fiber polarization controller. 

\begin{figure}
    \centering
    \includegraphics[width=1.0\linewidth]{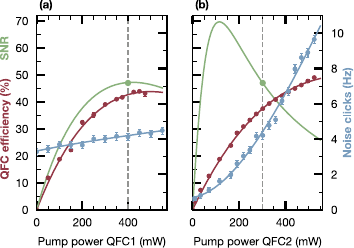}
    \caption{Characterization of the QFCs. Measurement of the external conversion efficiency of QFC1 (a) and QFC2 (b) - in red - and noise at the end of the communication link when changing (a) the pump power of QFC1 (power of QFC2 is fixed at 300\,mW) and (b) the pump power of QFC2 (power of QFC1 is fixed at 400\,mW)  - in blue. The ratio between the efficiency and the noise clicks is proportional to the SNR, which is depicted in green. All solid lines are obtained from fits to experimental data using equations that are found in~\cite{suppMat}. The vertical dashed grey lines indicate the pump powers used in the entanglement experiment for QFC1 and QFC2 respectively.}
    \label{fig:noise}
\end{figure}

The performance of the frequency conversion stages is determined by factors such as polarization preservation, conversion efficiency, and noise. The frequency conversion is achieved using a periodically-poled lithium niobate crystal in a waveguide. Each waveguide is placed inside a Sagnac interferometer, where the orthogonally polarized components are propagating in opposite directions. Adding a multichromatic waveplate enables coherent conversion of both counterpropagating polarizations without the need of any phase stabilization~\cite{Ikuta2018,VanLeent2020}. Fig.~\ref{fig:noise} (red curves) shows the external conversion efficiency (from input fiber to output fiber) as a function of the respective pump powers, indicating the need for $>$400 mW to reach the maximum conversion efficiencies of 44\% and 50\% for QFC1 and QFC2, respectively. This also leads to the generation of noise photons, in particular a broadband background due to anti-Stokes Raman scattering. Of special concern is the noise generated near the target telecom single photon wavelength, as it can significantly degrade entanglement fidelity if not properly mitigated. A minimum in the Raman noise spectrum was found at -394\,$\mathrm{cm}^{-1}$ (see Fig. S5 in \cite{suppMat}) which results in the wavelength of 1610\,nm for the pump laser and 1514\,nm for the telecom single photons.  To further mitigate the noise, a dedicated filtering stage is used. That includes a volume Bragg grating (VBG) with a full width at half maximum (FWHM) of 50\,GHz and a narrowband filter cavity with a FWHM of 20\,MHz centered at 780\,nm to maximize the entanglement fidelity of the atom-photon states.

Another key consideration is that the power-dependent anti-Stokes Raman noise at 1514\,nm can be upconverted to 780\,nm via the subsequent sum-frequency generation processes occurring in both QFCs. As shown by the blue curve in Fig.~\ref{fig:noise}, this behavior leads to distinct noise dynamics in the two conversion stages~\cite{Maring2018}. The curve displays the noise generated at each QFC stage, measured after the filtering system. In QFC1, Raman noise at 1514\,nm is depleted by this upconversion process and thus exhibits a sub-linear dependence on its pump power. For this measurement, the pump laser of QFC2 was operated at 300\,mW, resulting in a persistent noise offset of 3.3 Hz. Moreover, the noise originating from QFC1 propagates through the deployed optical fiber, where it undergoes attenuation, ultimately contributing less than 1\,Hz to the total noise rate. In stark contrast, in QFC2, newly generated Raman noise at 1514\,nm is directly converted to the target wavelength of 780\,nm. As a result, higher pump powers in QFC2 both generate more noise \textit{and} increase the efficiency of its conversion, leading to a super-linear dependence on the pump power. In the case of QFC1, the optimal SNR almost coincides with the maximum conversion efficiency. However, for QFC2, the noise increases rapidly with pump power, requiring a careful trade-off between maintaining a high SNR (see data point on the green curve in  Fig.~\ref{fig:noise}) and achieving sufficient conversion efficiency and hence an acceptable rate. For this, the pump powers have been set to 400\,mW for QFC1 and to 300\,mW for QFC2, respectively, resulting in a total noise rate of $4.2\pm0.1$\,Hz. To further maximize the SNR, entangled photons are detected within a fixed 60-ns window, resulting in an SNR of $47\pm8$.

\begin{figure}
\includegraphics[width=1.0\linewidth]{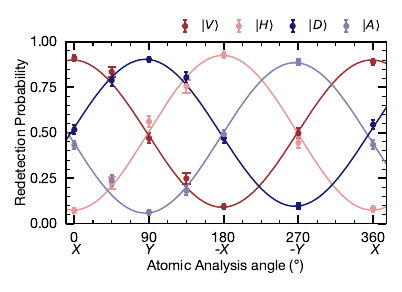}
\caption{Atom-photon state correlation at the telecom wavelength after 24\,km of fiber. The conditional redetection probability for different readout angles of the atom depending on the measured polarization state of the photon  (horizontal and vertical {($|H\rangle$, $|V\rangle$}) or diagonal and antidiagonal linear polarization ({$|D\rangle$, $|A\rangle$}) ) as indicated in the legend shows fringes with the average visibility $\bar V=82.4\pm0.8\%$. The measurement points at 45° and 135° are used for calculating the S-parameter, with result $S=2.30\pm0.08$, clearly violating Bell's inequality. Error bars represent one standard deviation, and the photon-detection window was set to 60\,ns.}
\label{fig:fringes} 
\end{figure}

To determine the quality of entanglement distribution over the deployed link, we first performed an intermediate measurement, i.e. observing the atom-photon entanglement before the back conversion. For this step, a filtering system consisting of a VBG with a FWHM of 26 GHz, and a narrowband filter cavity with a FWHM of 27 MHz centered at 1514 nm is used. Fig.~\ref{fig:fringes} confirms the atom-photon entanglement right before the back-conversion step. It shows the probability that the atom is not photoionized by the state-dependent photoionization measurement beams used for state detection (redetection probability), once the 1514\,nm photonic qubit was projected into a particular polarization state (indicated in the legend). The results are shown as a function of photoionization beam polarization angle, which selects the atomic qubit measurement basis. These data were obtained for a typical rate of 3.1 events per minute and 100\,min measurement time per data point ~\cite{suppMat}. We observe high visibility for both the horizontal-vertical, $V_{HV}=81.7\pm0.7\%$, and diagonal-antidiagonal linear polarization of the photon, $V_{DA}=83.0\pm1.6\%$. Averaging over the different bases and assuming white noise, the mean visibility $\bar{V} = 82.4\pm0.8\%$ is related to the fidelity $F$ as~\cite{Zhou2024}

\begin{equation}
\mathcal{F}\geq\frac{1}{4}+\frac{3}{4}\bar{V}
\label{eq:fidelity}
\end{equation}

\noindent which leads to a fidelity bound of $\mathcal{F}\geq 86.8\pm 0.6\%$, a marginal drop compared to the fidelity of $\mathcal{F}\geq 87.6\pm 2.0\%$ for 0\,km. 

Using the values in Fig.~\ref{fig:fringes} taken at atomic analysis angles of 45° and 135°, one obtains a Bell parameter of $S=2.30 \pm 0.08$, violating the Bell inequality by more than 3 standard deviations. 

\begin{figure}
\includegraphics[width=1.0\linewidth]{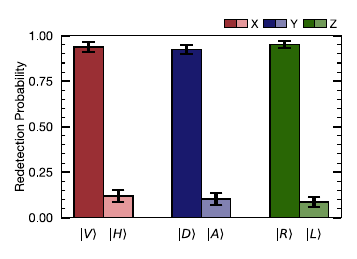}
\caption{Atom-photon state correlation at 780\,nm after 24\,km of fiber. Conditional redetection probability of the atom at maximum correlation points measured in three different basis combinations for both the atom (X,Y and Z) and the photon measurement. The horizontal axis indicates the photon qubit polarization measurement result. The different values obtained for the VH, DA and LR bases lead to an average visibility of $\bar V=82.5\pm2.0\%$. The photon detection rate for this measurement was 0.7 min$^{-1}$. Error bars represent one standard deviation.}
\label{fig:max-corr}
\end{figure}

Finally, we characterize the entanglement distribution after the photonic qubit is back-converted to 780\,nm at MPQ. Fig.~\ref{fig:max-corr} shows the atom redetection probability, this time conditioned on the 780\,nm photonic qubit projection into a particular polarization state. For all the measurements shown in this figure, the atomic qubit measurement basis is set at different orthogonal angles for which maximum atom-photon state correlation is expected. We observe 0.7 events per minute for a total measurement time of 630 minutes and obtain $\mathcal{F}\geq 86.9 \pm 1.5\%$, which is significantly higher than the upper bound for non-entangled states $\mathcal{F}\leq 0.5$~\cite{Sackett2000}. The lower SNR of  $47\pm8$ compared to the case without link reduces the fidelity only marginally~\cite{suppMat}. The overall entanglement fidelity is limited by imperfect atomic state manipulation (4.9\%), readout (5.5\%) and decoherence (2.4\%), and photon qubit decoherence along the propagation path ($0.2\%$ right after polarization stabilization \cite{suppMat}). 
Furthermore, by measuring at additional angles we obtain a Bell parameter of $S=2.29 \pm 0.08$, again clearly violating the Bell inequality and showing the preservation of the entanglement after the back-conversion step. This certifies the distribution of atom-photon entanglement in the visible band over the long-distance communication link.

While these results clearly show the distribution of atom-photon entanglement over a metropolitan link, future experiments and applications may require the optimization of several parameters. One significant limitation of the entanglement distribution rate is the single photon fiber collection efficiency (1\%), which could be significantly improved by coupling the atom to an optical cavity. In addition, generating multiplexed atom-photon entanglement using atomic arrays coupled to a cavity would further enhance the rate~\cite{Hartung2024, Canteri2025}. Depending on the number of atoms in the array these two measures could bring a rate improvement of more than two orders of magnitude. Enhancing the single photon fiber collection efficiency would also improve the photon measurement SNR, and would enhance the entangled state fidelity. A higher SNR would furthermore relax the current trade-off between conversion efficiency and noise in QFC2, enabling operation at higher pump power and thus increased conversion efficiency. Another strategy to improve the SNR is to decrease the photon noise, mostly generated during the quantum frequency conversion processes. Cooling the nonlinear crystal waveguide is one way to reduce anti-Stokes Raman scattering noise~\cite{Kuo2018}. Alternatively, converting to the O-Band or  performing two-stage quantum frequency conversion at the end of the link would also allow to operate in a wavelength regime in which Raman scattering noise is minimized, although at the cost of having additional signal losses~\cite{Luo2022, Schafer2025}.

In conclusion, we have demonstrated a metropolitan quantum communication link distributing entanglement between an atom and a photon that undergoes wavelength conversion to the telecom band and back to the near-visible. The first conversion step enables low‑loss transmission through the 24\,km fiber, while the final telecom‑to‑visible conversion enables subsequent quantum‑network operations with high fidelity. One future task would be to map the state of the distributed photon to another atomic quantum information processing node in a heralded manner~\cite{Brekenfeld2020}. Remote and heralded atomic entanglement can then be used for instance to perform advanced quantum communication protocols such as device-independent quantum key distribution~\cite{Pironio2009},  repeaters~\cite{Briegel1998, Langenfeld2021} or teleportation~\cite{Langenfeld2021_Teleportation}. Furthermore, the near-visible wavelength of the distributed photon is also ideal to connect with nondestructive photon qubit tracking systems, which allows to mitigate the photon loss problem in subsequent teleportation schemes and loss-sensitive qubit measurements~\cite{Niemietz2021}. Finally, the telecom-to-visible conversion also makes it possible to combine fiber transmission directly with free-space communication links using telescopes and satellites, benefiting from low atmospheric absorption and beam divergence~\cite{Liao2017,Abasifard2024}. Overall, our results mark a significant step towards integrating remote quantum devices with existing fiber networks, quantum information processing protocols and even free-space applications for long-distance quantum communication.

\begin{acknowledgments}
This work was funded by the Deutsche Forschungsgemeinschaft (German Research Foundation) under Germany’s Excellence Strategy – EXC-2111 – 390814868, by the German Federal Ministry of Research, Technology and Space (Bundesministerium für Forschung, Technik und Raumfahrt, BMFTR) through projects QR.X (16KISQ002, 16KISQ019), QR.N (16KIS2187, 16KIS2189), and  QuKuK (16KIS1621), as well as by the Munich Quantum Valley lighthouse project NeQuS (Z.5-F5121.17.1/5/80) under the Hightech Agenda Bayern Plus of the Bavarian state government. 
\end{acknowledgments}

\section*{Data availability}
The data that support the findings of this article are openly available at \cite{DataZenodo}, embargo periods may apply.

\bibliography{bibliography}%

\end{document}